\documentclass[twocolumn, prl, superscriptaddress, longbibliography]{revtex4-1}
\usepackage{verbatim,amsmath,amssymb,graphicx}
\usepackage{hyperref}
\hypersetup{
        colorlinks=true,       
        linkcolor=red,        
        citecolor=blue,     
        urlcolor=blue, pdfauthor={some author},pdftitle={eye-catching title}}
\renewcommand{\section}[1]{{\par\it #1.---}\ignorespaces}

\begin{document}
\title{Floquet engineering to reactivate a dissipative quantum battery}
\author{Si-Yuan Bai}
\affiliation{School of Physical Science and Technology, Lanzhou University, Lanzhou 730000, China}
\author{Jun-Hong An}
\email{anjhong@lzu.edu.cn}
\affiliation{School of Physical Science and Technology, Lanzhou University, Lanzhou 730000, China}

\begin{abstract}
As an energy storing and converting device near atomic size, a quantum battery (QB) promises enhanced charging power and extractable work using quantum resources. However, the ubiquitous decoherence causes its cyclic charging-storing-discharging process to become deactivated, which is called aging of the QB. Here, we propose a mechanism to overcome the aging of a QB. It is found that the decoherence of the QB is suppressed when two Floquet bound states (FBSs) are formed in the quasienergy spectrum of the total system consisting of the QB-charger setup and their respective environments. As long as either the quasienergies of the two FBSs are degenerate or the QB-charger coupling is large in the presence of two FBSs, the QB exposed to the dissipative environments returns to its near-ideal cyclic stage. Our result supplies an insightful guideline to realize the QB in practice using Floquet engineering.

\end{abstract}
\maketitle

\section{\label{sec:level1}Introduction}
A battery is a device that stores chemical energy and converts it into electrical energy. The development of microscopic electronic equipment appeals to batteries of molecular or even atomic size, where quantum mechanics takes effect. It inspired the birth of a quantum battery (QB) \cite{PhysRevE.87.042123}. QBs hold the promise of higher energy storing density in large-scale integration and faster charging power than its classical counterpart \cite{Popescu2014}. Studies in past years have shown the distinguished role of quantum resources in improving the performance of QBs, such as work extraction \cite{PhysRevLett.122.047702,PhysRevLett.111.240401,Friis2018precisionwork}, charging power \cite{PhysRevLett.118.150601,PhysRevLett.120.117702,PhysRevE.99.052106,Binder_2015,Chen2020}, and stabilization of stored energy \cite{PhysRevResearch.2.013095}. This progress paves the way to realize QBs from the physical principles.

However, the performance of QBs well developed in unitary evolution \cite{PhysRevE.87.042123,PhysRevA.97.022106,Binder_2015,PhysRevLett.118.150601,PhysRevE.100.032107,PhysRevResearch.2.023113,PhysRevE.101.062114,PhysRevLett.120.117702,Crescente_2020} is obscured by the ubiquitous decoherence in practice. It severely constrains the practical realization of QBs. Degrading the quantum resources of the QB, decoherence caused by the inevitable interactions of the QB with its environment generally deactivates the QB, which is called the aging of the QB \cite{PhysRevA.100.043833}. Thus, the general analysis of QBs must resort to the open system approach. Previous studies on this topic are mostly based on Markovian approximation \cite{PhysRevA.100.043833,PhysRevLett.122.210601,PhysRevB.99.035421,PhysRevLett.125.040601,JJLiu2019,PhysRevApplied.14.024092}. It has been found that the efficiency of QBs is reduced with time in the Markovian approximation description to decoherence \cite{PhysRevA.100.043833,PhysRevLett.122.210601,PhysRevB.99.035421}. Based on the fact that the Markovian approximation may miss physics, especially when systems and environments form a hybrid bound state \cite{PhysRevA.81.052330,PhysRevLett.109.170402,PhysRevB.95.161408,PhysRevResearch.1.023027,PhysRevLett.123.040402}, the non-Markovian description of decoherence \cite{RevModPhys.88.021002,Rivas_2014,LI20181} in QBs is much desired. The recent studies indeed reveal the constructive role of the non-Markovian effect in improving the performance of QBs \cite{Uzdin2016,Kamin_2020,Carrega2020}. However, they did not touch on the cyclic process of charging, storing, and discharging of QBs, which causes the QB system to be periodically dependent on time. Therefore, a complete analysis of the decoherence dynamics of the cyclic charing-storing-discharging process of QB and an efficient method to postpone the aging of QBs are still absent.

We here propose a mechanism to overcome the aging of a dissipative QB. By modeling the QB and the charger as two-level systems, we investigate the cyclic charing-storing-discharging evolution of the QB-charger setup exposed to dissipative environments by Floquet theory. It is found that, in sharp contrast to the Markovian result, the energy of the QB in the non-Markovian dynamics exhibits diverse long-time features, including complete decay, energy trapping, and persistent oscillation. Our analysis demonstrates that they are essentially determined by the different numbers of Floquet bound states (FBSs) formed in the quasienergy spectrum of the total system consisting of the QB-charger setup and their respective environments. It gives us insightful instruction to manipulate the quasienergies of the two FBSs such that the QB is reactivated to its near-ideal cyclic stage. This is realizable when the quasienergies of the two FBSs are degenerate or the QB-charger coupling is large. Our result may provide a guideline to realize QBs in practice.

\begin{figure}
\includegraphics[width=.46\columnwidth]{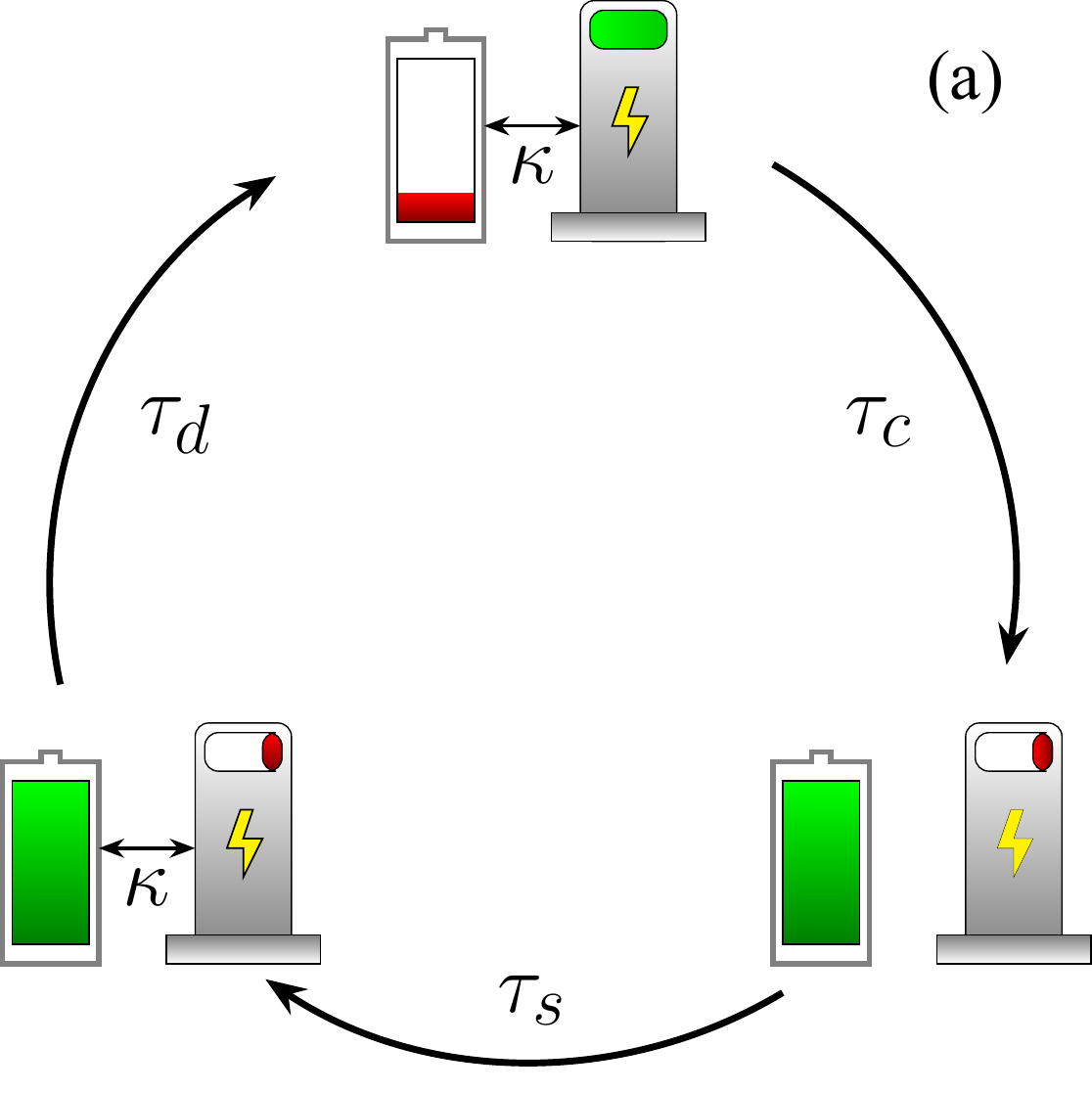}\includegraphics[width=.48\columnwidth]{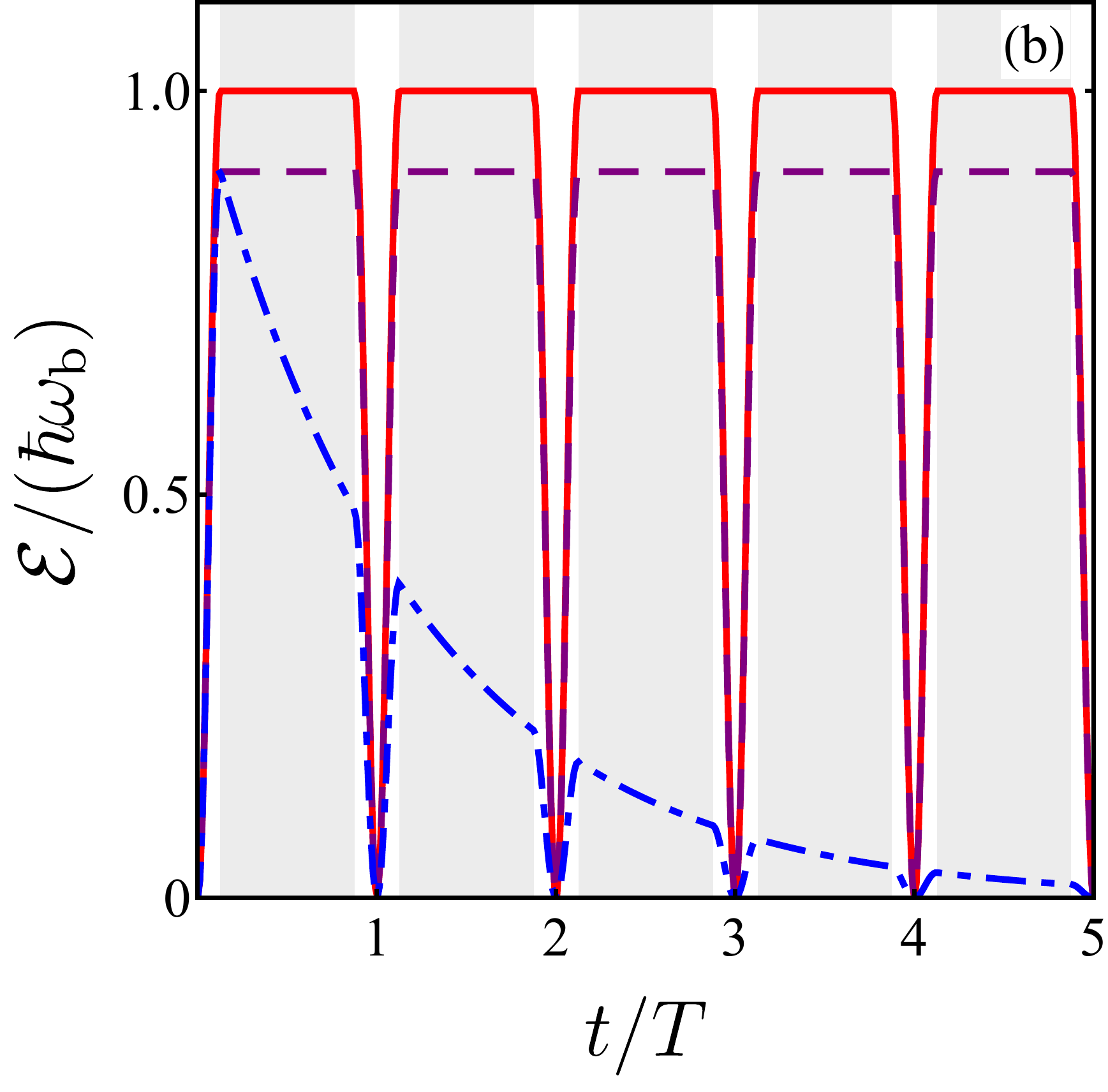}
\caption{ (a) Schematic illustration of the QB-charger setup, where $T=\tau_c+\tau_s+\tau_d$ is the one-cycle period with $\tau_c$, $\tau_s$, and $\tau_d$ being the time of charging, storing, and discharging. (b) Evolution of the QB energy $\mathcal{E}(t)$ when $\delta=0$ (red solid line) and $10 \omega_\text{b}$ (purple dashed line) in the ideal case. The blue dot-dashed line is the result in the presence of the Markovian decoherence with $\Gamma=0.5\omega_\text{b}$ and $\delta=0$. We use $\kappa=15 \omega_\text{b}$, $\tau_s=2\pi/(10 \omega_\text{b})$ and $\tau_c=\tau_d=\pi/(2\kappa)$ .}\label{idesch}
\end{figure}

\section{Ideal QB scheme\label{sec:system}}
A battery is a device that stores and converts energy. The basic idea of a QB is to use the discrete energy levels of a quantum system for energy storage and conversion \cite{Gumberidze2019,Allahverdyan_2004,Francica2017,PhysRevB.98.205423,arXiv:1805.05507,PhysRevB.100.115142,PhysRevB.99.205437}. Without loss of generality, we consider a two-level system as the QB \cite{Carrega2020,Kamin_2020,PhysRevLett.122.047702,PhysRevLett.120.117702}. Charging the QB is to change its state from the ground state to the excited state. This can be realized by coupling the QB to a quantum charger, which also can be modeled by a two-level system. After the charging, the coupling is switched off and the energy is well stored in the QB. The usage of the QB causes its discharging, which is described by its interaction with a target system. We here describe the discharging by switching on the QB-charger coupling again for simplicity. A good performance of the QB means that the charging, storing, and discharging processes works cyclically [see Fig. \ref{idesch}(a)]. The cyclic evolution is governed by the Hamiltonian
\begin{equation}
\hat{H}_0(t)=\sum_{l=\text{b,c}}\hbar\omega_{l}\hat{\sigma}_{l}^{\dagger}\hat{\sigma}_{l}+\hbar\kappa f(t)(\hat{\sigma}_\text{b}^{\dagger}\hat{\sigma}_\text{c}+\text{H.c.}),\label{eq:Ham-sys}
\end{equation}where $\hat{\sigma}_l=|g_l\rangle\langle e_l|$, with $|g\rangle$ and $|e\rangle$ being the ground and excited states, are the transition operators of the QB and charger with frequency $\omega_l$, and $\kappa$ is their coupling strength. The time-dependent $f(t)$ is
\begin{equation}
f(t)=\begin{cases}
1, & nT<t\le\tau_\text{c}+nT\\
0, & \tau_{c}+nT<t\le\tau_{c}+\tau_{s}+nT\\
1, & \tau_{c}+\tau_{s}+nT<t\le(n+1)T
\end{cases}\label{eq:f(t)}, ~n\in \mathbb{N},
\end{equation}
where $\tau_c$, $\tau_s$, and $\tau_d$ are the time of charging, storing, and discharging, respectively, and $T=\tau_c+\tau_s+\tau_d$ is the period. For the initial state $|\psi(0)\rangle=|g_\text{b},e_\text{c}\rangle$, we obtain the evolved state $|\psi(t)\rangle$ governed by Eq. \eqref{eq:Ham-sys}. The performance of the QB is quantified by its energy $\mathcal{E}(t)=\hbar\omega_\text{b}\langle\psi(t)|\hat{\sigma}_\text{b}^{\dagger}\hat{\sigma}_\text{b}|\psi(t)\rangle$. If $\mathcal{E}(t)$ reaches the maximum $\hbar\omega_\text{b}\kappa^2/\Omega^2$ with $\Omega=\sqrt{\kappa^2+\delta^2}$ and $\delta=(\omega_\text{c}-\omega_\text{b})/2$ at the end of each charging step, and empties its energy at the end of each discharging step, then the QB works in an ideal stage. This is achieved when $\Omega\tau_c=({1/2}+n_1)\pi$, $\delta\tau_s=\pi n_2$, and $\Omega\tau_d=({1/2}+n_3)\pi$ with $n_j\in\mathbb{N}$ \cite{SMP}. Further under the resonant condition $\delta=0$, $\mathcal{E}(t)$ reaches its optimal value $\hbar\omega_\text{b}$. We show in Fig. \ref{idesch}(b) the evolution of $\mathcal{E}(t)$. It is zero at the initial time of each charging-storing-discharging cycle and reaches its maximum after the charging process $\tau_c$. After a storage in the duration $\tau_s$, the energy is emptied after the discharging process.

\section{Effect of dissipative environments\label{sec:withdissipation}}
Actually, any realistic QB-charger setup is inevitably influenced by its environments and experiences decoherence. It is important to access the performance of
the QB when the deocoherence of both the QB and the charger is considered. Determined by whether the system has energy exchange with the environment, the decoherence can be classified into dephasing and dissipation. The dephasing arises from elastic collisions in a dense atomic ensemble or elastic phonon scattering in a solid system, neither of which are significant in our single two-level system scenario \cite{Carmichael}. Thus, we focus on the impact of the dissipative environments on the QB, whose Hamiltonian is $\hat{H}(t)=\hat{H}_0(t)+\hat{H}_\text{E}+\hat{H}_\text{I}$ with
\begin{eqnarray}
\hat{H}_\text{E}=\sum_{l,\mathbf{k}}\hbar\omega_{l,\mathbf{k}}\hat{b}_{l,\mathbf{k}}^{\dagger}\hat{b}_{l,\mathbf{k}},
~\hat{H}_\text{I}=\sum_{l,\mathbf{k}}\hbar(g_{l,\mathbf{k}}\hat{b}_{l,\mathbf{k}}^{\dagger}\hat{\sigma}_l+\text{H.c.}),~\label{eq:H_interaction-1}
\end{eqnarray}
where $\hat{b}_{l,\textbf{k}}$ are the annihilation operators of the ${\bf k}$th mode with frequency $\omega_{l,\mathbf{k}}$ of the environments felt by the QB and charger, and $g_{l,\mathbf{k}}$ are their coupling strengths.

The dynamics is calculated by expanding the state as
\begin{equation}
|\Psi(t)\rangle=\sum_{l=\text{b,c}}\big[u_l(t)\hat{\sigma}_l^\dag +\sum_{\mathbf{k}}\eta_{l,\mathbf{k}}(t)\hat{b}^\dag_{l,\mathbf{k}}\big]|{\text{\O}}\rangle,\label{eq:state}
\end{equation}
where $|\text{\O}\rangle\equiv|g_\text{b},g_\text{c},\{0_\mathbf{k}\}_\text{b},\{0_\mathbf{k}\}_\text{c}\rangle$, $u_\text{b}(0)=\eta_{l,\mathbf{k}}(0)=0$ and $u_\text{c}(0)=1$. We derive from the Schr\"{o}dinger equation governed by $\hat{H}(t)$ that the coefficient $u_l(t)$ satisfies
\begin{equation}
\dot{u}_{l}(t)+i\omega_{l}u_{l}(t)+i\kappa f(t)u_{l'}(t)+\int_{0}^{t}\nu_l(t-\tau)u_{l}(\tau)d\tau =0\label{eq:dynamics}
\end{equation}
where $l\ne l'$ and the correlation function $\nu_l(x)=\int_0^\infty J_l(\omega)e^{-i\omega x}d\omega$ with $J_l(\omega)=\sum_{\bf k}|g_{l,{\bf k}}|^2\delta(\omega-\omega_{l,{\bf k}})$ being the environmental spectral densities. The energy of the QB in this decoherence case reads $\mathcal{E}(t)=\hbar\omega_\text{b}|u_\text{b} (t)|^{2}$.

We first consider the situation of $\omega_\text{b}=\omega_\text{c}\equiv\omega_0$ and $J_\text{b}(\omega)=J_\text{c}(\omega)\equiv J(\omega)$. Then Eqs. \eqref{eq:dynamics}
are decoupled into $\dot{v}_{\pm}(t)\pm i\kappa f(t)v_{\pm}(t)+\int_{0}^{t}\nu'(t-\tau)v_{\pm}(\tau)d\tau=0$ with $\nu'(x)=\nu(x)e^{i\omega_0 x}$ by defining $v_{\pm}(t)=[u_\text{c} (t)\pm u_\text{b} (t)]e^{i\omega_{0}t}$. When the QB/charger-environment coupling is weak and the environmental correlation time is smaller than those of the QB/charger, we apply the Markovian approximation via replacing $v_{\pm}(\tau)$ by $v_{\pm}(t)$ and extending the upper limit of the time integration to infinity. Then we obtain $v_{\pm}(t)=e^{-(\Gamma+i\Delta)t\mp i\kappa\int_{0}^{t}f(\tau)d\tau}$, where $\Gamma=\pi J(\omega_{0})$ and $\Delta=\mathcal{P}\int d\omega\frac{J(\omega)}{\omega_0-\omega}$ with $\mathcal{P}$ being the principal integral \cite{PhysRevE.90.022122}. They induce $u_\text{b}(t)=-i\sin[\kappa\int_{0}^{t}f(\tau)d\tau]e^{-[\Gamma+i(\omega_{0}+\Delta)]t}$ and the Markovian approximate energy as
\begin{equation}
\mathcal{E}_\text{M}(t)=\hbar\omega_0 e^{-2\Gamma t}\sin^{2}\big[\kappa\int_{0}^{t}f(\tau)d\tau\big].\label{eq:markov}
\end{equation} We see that $\mathcal{E}_\text{M}(t)$ tends to zero [see the dashed blue line in Fig. \ref{idesch}(b)] and the QB is deactivated by the Markovian decoherence. It is called the aging of the QB \cite{PhysRevA.100.043833,Kamin_2020,Carrega2020}.
\begin{figure}[tbp]
\includegraphics[width=.96\columnwidth]{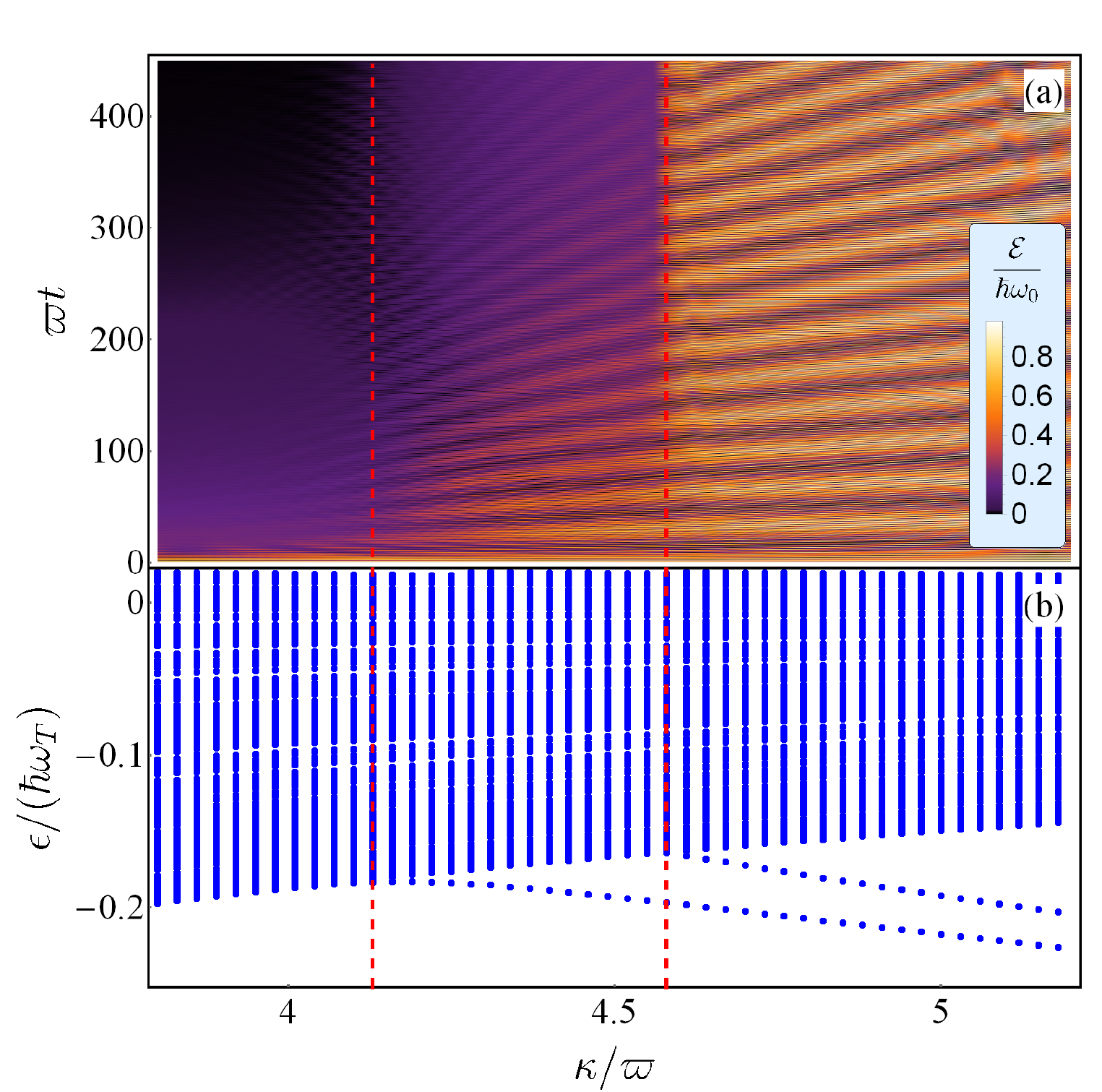}
\caption{(a) Evolution of $\mathcal{E}(t)$ and (b) quasienergy spectrum of the total system in different coupling strength $\kappa$. We use $N=30$, $\omega_0=2\varpi$, $g=q=0.5\varpi$, and $\tau_c=\tau_s=\tau_d=\pi/(2\kappa)$.  }\label{fig:case1}
\end{figure}

In the non-Markovian case, the environmental spectral densities are needed. We consider that each environment is described by $N\times N$ bosonic modes on a square lattice. The modes have an identical frequency $\varpi$ and nearest-neighbor coupling strength $\hbar q$ \cite{PhysRevLett.119.143602,PhysRevLett.122.203603,Su-Peng2019}. The QB and charger reside in the site $\mathbf{n}=(0,0)$ and couple to the mode of this site in strength $\hbar g$. In the momentum space, the environmental Hamiltonian is written as Eqs. \eqref{eq:H_interaction-1} with $\omega_{l,\mathbf{k}}=\varpi-2q(\cos k_{x}+\cos k_{y})$ and $g_{l,\mathbf{k}}= g/N$. The spectral density is calculated as \cite{PhysRevA.96.043811}
\begin{equation}
J(\omega)=\frac{g^{2}}{2q\pi^{2}}\Theta(4q-|\omega-\varpi|)K\Big(1-\frac{(\omega-\varpi)^{2}}{16q^{2}}\Big),
\end{equation}
where $\Theta(x)$ is the step function and $K(x)$ is the complete elliptic integral of the first kind. The energy $\mathcal{E}(t)$ is obtained by numerically solving Eqs. \eqref{eq:dynamics}. We plot $\mathcal{E}(t)$ in different coupling strength $\kappa$ in Fig. \ref{fig:case1}(a). In contrast to the Markovian result, three distinct regions where $\mathcal{E}(t)$ shows qualitatively different dynamics are observed in this non-Markovian case. When $\kappa\lesssim 4.1\varpi$, $\mathcal{E}(t)$ tends to zero and the cyclic evolution is destroyed, which is consistent with the previous results \cite{PhysRevA.100.043833,Kamin_2020,Carrega2020}. When $4.1\varpi\lesssim\kappa\lesssim 4.6\varpi$, $\mathcal{E}(t)$ approaches the finite values with tiny-amplitude oscillation, where the energy is partially trapped and the cyclic evolution still does not work. When $\kappa\gtrsim 4.6\varpi$, $\mathcal{E}(t)$ approach a Rabi-like persistent oscillation with multiple frequencies. The two latter cases signify that the dissipation of the QB is suppressed.

\section{Floquet engineering to reactivate the QB}\label{sec:Results}
The diverse non-Markovian dynamics can be explained by the Floquet theory \cite{PhysRev.138.B979,PhysRevA.7.2203}, which supplies us an insightful understanding of the temporally periodic system $\hat{H}(t)=\hat{H}(t+T)$ \citep{doi:10.1063/1.5144779,PhysRevA.91.052122,PhysRevLett.121.080401}. According to the theory, there are a set of time-periodic basis $|\phi_{\alpha}(t)\rangle=|\phi_{\alpha}(t+T)\rangle$ determined by the Floquet equation
$[\hat{H}(t)-i\hbar\partial_{t}]|\phi_{\alpha}(t)\rangle=\epsilon_{\alpha}|\phi_{\alpha}(t)\rangle$ such that any state evolves as $|\Psi(t)=\sum_{\alpha}c_{\alpha}e^{ \frac{-i}{\hbar} \epsilon_{\alpha}t}|\phi_{\alpha}(t)\rangle$ with $c_\alpha=\langle \phi_{\alpha}(0)|\Psi(0)\rangle$.
The time independence of $\epsilon_\alpha$ and $c_\alpha$ implies that $\epsilon_\alpha$ and $|\phi_{\alpha}(t)\rangle$ play the same role as eigenenergies and stationary states of a static system. Thus they are called quasienergies and quasistationary states, respectively. Because $e^{in\omega_T t}|\phi_{\alpha} (t)\rangle$ with $\omega_T=2\pi/T$ is also the solution of the Floquet equation with eigenvalues $\epsilon_\alpha+n\hbar\omega_T$, the quasienergies are periodic with period $\hbar\omega_T$ and one generally chooses them within $(-\hbar\omega_T/2,\hbar\omega_T/2]$ called the first
Brillouin zone. Actually the Floquet equation is equivalent to $\hat{U}_T|\phi_{\alpha}(0)\rangle=e^{-i\epsilon_\alpha T/\hbar}|\phi_{\alpha}(0)\rangle$ with $\hat{U}_T$ the one-period evolution operator, from which $\epsilon_\alpha$ and $|\phi_{\alpha}(0)\rangle$ is obtainable. Then applying the arbitrary-time evolution operator $\hat{U}_t$ on $|\phi_{\alpha}(0)\rangle$, $|\phi_{\alpha}(t)\rangle$ are obtained \cite{Eckardt_2015}.

\begin{figure}
\includegraphics[width=.96\columnwidth]{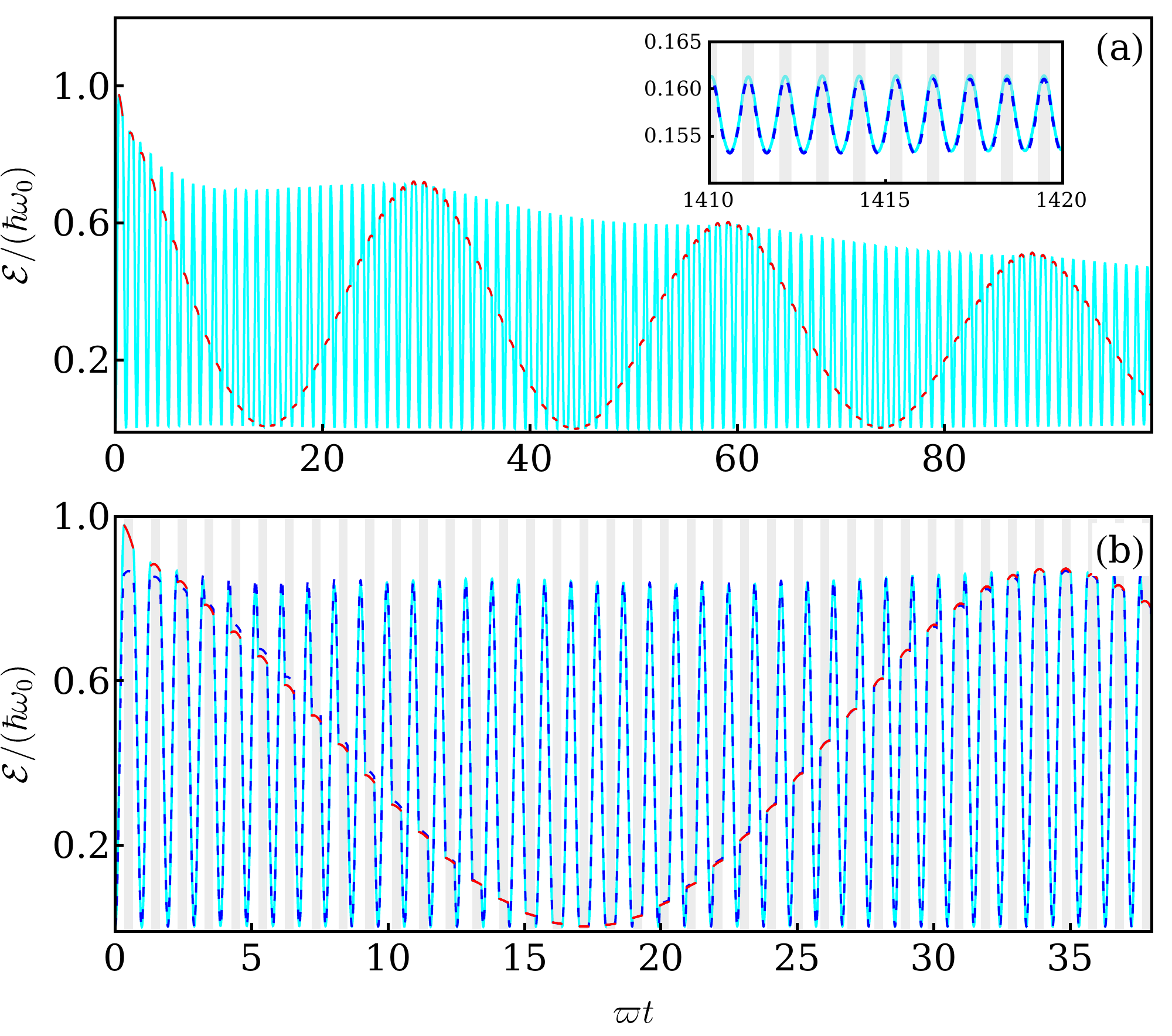}
\caption{Evolution of $\mathcal{E}(t)$ in the presence of one (a) and two (b) FBSs. The cyan solid lines are the numerical results by solving Eq. \eqref{eq:dynamics}. The blue dashed lines are evaluated from Eq. \eqref{stdaval}. The inset of (a) shows the long-time behavior. The energy in the storage time duration marked by the gray area is highlighted by red segments. Parameters are the same as Fig. \ref{fig:case1} except  $\kappa=4.5\varpi$ in (a) and $4.8\varpi$ in (b).} \label{comprs}
\end{figure}

The quasienergy spectrum in Fig. \ref{fig:case1}(b) shows that besides a continuous quasienergy band, isolated quasienergy levels in the band-gap area are present. We call such isolated levels FBSs, which play important roles in many systems \cite{PhysRevA.99.010102,PhysRevB.100.115411,PhysRevB.93.245434} and have been experimentally observed \cite{Kri2018,PhysRevLett.108.043603,Liu2017,Hood201603788}. The two branches of FBSs divide the spectrum into three regions: without FBS when $\kappa\lesssim 4.1\varpi$, one FBS when $4.1\varpi\lesssim\kappa\lesssim 4.6\varpi$, and two FBSs when $\kappa\gtrsim 4.6\varpi$. It is interesting to find that the regions match well with those where $\mathcal{E}(t)$ shows different behaviors, i.e., complete decay, energy trapping, and persistent oscillation in Fig. \ref{fig:case1}(a). The similar correspondence between the bound states and the dynamics has been reported before \cite{PhysRevA.97.023808,PhysRevResearch.1.023027}.  To understand it, we, according to the Floquet theory, rewrite Eq. \eqref{eq:state} as
\begin{equation}
|\Psi(t)\rangle=\sum_{j=1}^Mc_je^{{-i\over\hbar} \epsilon_{0j}t}|\phi_{0j}(t)\rangle+\sum_{\beta\in\text{CB}}d_{\beta}e^{{-i\over\hbar}\epsilon_{\beta}t}|\phi_{\beta}(t)\rangle, \label{flqdt}
\end{equation}
where $M$ is the number of FBSs, $c_j\equiv\langle \phi_{0j}(0)|\Psi(0)\rangle$, and $d_{\beta}\equiv\langle \phi_\beta(0)|\Psi(0)\rangle$. Due to the out-of-phase interference of different terms in continuous energy $\epsilon_\beta$, the contribution of the second term in Eq. (\ref{flqdt}) to $\mathcal{E}(t)$ approaches zero in the long-time condition. Thus $\mathcal{E}(\infty)$ only contains the contributions of the FBSs, i.e.,
\begin{eqnarray}
{\mathcal{E}(\infty)\over\hbar\omega_0}=\sum_{jj'=1}^Mc_jc_{j'}^*e^{{-i\over\hbar}(\epsilon_{0j}-\epsilon_{0j'})t}\langle \phi_{0j'}(t)|\hat{\sigma}_\text{b}^\dag\hat{\sigma}_\text{b}|\phi_{0j}(t)\rangle.~~~~\label{stdaval}
\end{eqnarray}
If the FBS is absent, then $M=0$ and $\mathcal{E}(\infty)=0$. If one FBS is formed, then $M=1$ and $\mathcal{E}(\infty)$ shows a perfect oscillation with the same frequency $\omega_T$ as the FBS. If two FBSs are formed, then $M=2$ and $\mathcal{E}(\infty)$ shows the persistent oscillation with multiple frequencies jointly determined by $\omega_T$ and $\Delta\epsilon_0\equiv|\epsilon_{01}-\epsilon_{02}|$. The validity of Eq. \eqref{stdaval} is confirmed by Fig. \ref{comprs}. Although the dissipation is efficiently suppressed for $M=1$, the QB cannot empty its energy at the end of each discharging step [see Fig. \ref{comprs}(a)]. The QB in this case is still badly performed. It is remarkable to find that the energy for $M=2$ almost behaves as perfectly as the ideal case in Fig. \ref{idesch}(b) except that the energy in each storage time duration oscillates in a frequency $\Delta\epsilon_0$ [see Fig. \ref{comprs}(b)] due to the interference of the two FBSs. If this oscillation is sufficiently avoided, then the QB would return to its ideal stage.

\begin{figure}
\includegraphics[width=1\columnwidth]{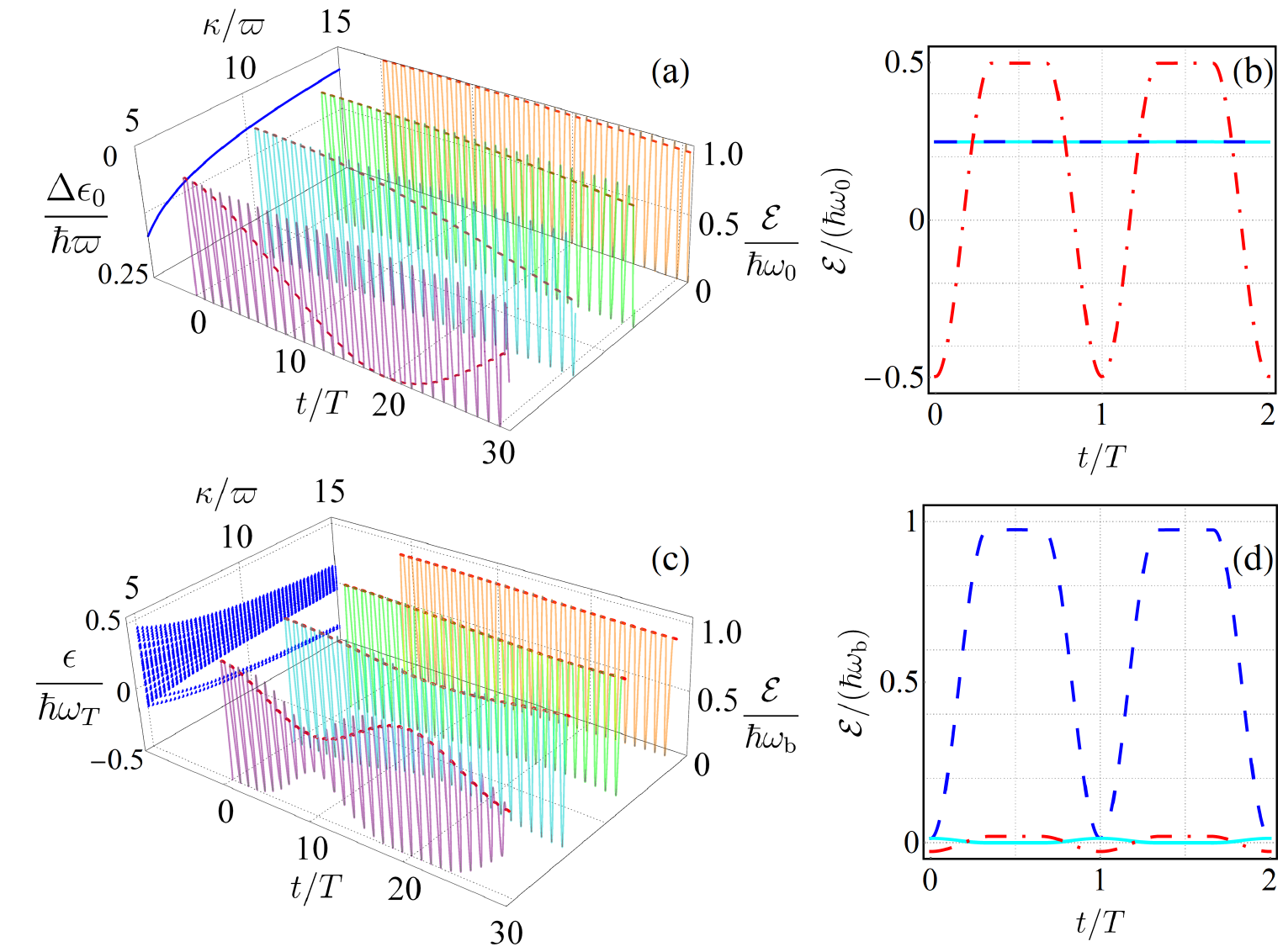}
\caption{(a) Evolution of $\mathcal{E}(t)$ and quasienergy difference $\Delta\epsilon_0$ (blue solid line) of the two FBSs in different $\kappa$ with $\delta=0$. Diagonal terms for $j=j'=1$ (cyan solid line) and $2$ (blue dashed line), and interference terms for $j\neq j'$ (red dotdashed line) of Eq. \eqref{stdaval} when $\kappa=15\varpi$, $\delta=0$ (b) and  $\kappa=15\varpi$, $\delta=0.5\varpi$ (d). (c) Evolution of $\mathcal{E}(t)$ and quasienergy spectrum $\epsilon$ in different $\kappa$ when $\delta=0.5\varpi$. The energy in storage time duration is highlighted by red segments in (a) and (c). The other parameter values are the same as Fig. \ref{comprs}.  }\label{stablid}
\end{figure}

With the mechanism of the dominated role of the FBSs in the dynamics of the QB at hand, we can reactivate the QB under the environmental influence by controlling the quasienergies of the two FBSs.  This can be realized when $\Delta \epsilon_0$ approaches zero. Using perturbation theory, we can evaluate $\Delta \epsilon_0=\hbar\sum_{\mathbf{k},n} \frac{(1-2n) g_{\mathbf{k}}^2|\tilde{y}_{n}|^2}{(\frac{1}{2}-n)^2\omega_T-(\omega_0-\omega_{\mathbf{k}})^2/\omega_T}$, where $\tilde{y}_n=\frac{1}{T} \int_0^Tdt e^{-i n\omega_T t}y(t)$ with $y(t)$ given in the Supplemental Material \cite{SMP}. Keeping only the leading term $n=0$, we can evaluate $\Delta \epsilon_0 \simeq \frac{3\hbar g^2|\tilde{y}_0|^2}{4\pi^2\kappa}$ in the large $\kappa$ condition. Figure \ref{stablid}(a) confirms that the energy oscillation is slowed down when $\Delta\epsilon_0$ tends to zero with the increase of $\kappa$. In the large $\kappa$ condition, the two diagonal terms of Eq. \eqref{stdaval} almost become a constant and the interference terms dominate the large-amplitude oscillation in period $T$ [see Fig. \ref{stablid}(b)]. The oscillation of the energy in the storing time duration is thus stabilized. The leading-order perturbation solutions of the two FBSs are $|\phi_{01}(t)\rangle\simeq y(t)e^{-i\omega_Tt}\frac{\hat{\sigma}^\dag_\text{b}+\hat{\sigma}^\dag_\text{c}}{\sqrt{2}}|{\O}\rangle$ and $|\phi_{02}(t)\rangle=y^*(t)\frac{\hat{\sigma}^\dag_\text{b}-\hat{\sigma}^\dag_\text{c}}{\sqrt{2}}|{\O}\rangle$ \cite{SMP}, which verify the evolution behaviors in Fig. \ref{stablid}(b).

We can extend our result to the nonresonant case. Although the quasienergy difference $\Delta \epsilon_0$ of the two FBSs cannot be zero anymore, we still have a chance to make the QB return to its near-ideal stage by increasing $\kappa$ [see Fig. \ref{stablid}(c)]. In the large $\kappa$ regime in the presence of a small detuning $\delta$, we can evaluate the leading-order perturbation solutions of the two FBSs as $|\phi_{01}(0)\rangle\simeq \hat{\sigma}^\dag_\text{b}|{\text{\O}}\rangle$ and $|\phi_{02}(0)\rangle\simeq \hat{\sigma}^\dag_\text{c}|{\text{\O}}\rangle$ \cite{SMP}. They are contained in the initial state $|\Psi(0)\rangle$ as components with probability amplitudes $c_1=0$ and $c_2=1$, respectively. Then according to Eq. \eqref{stdaval}, only the FBS $|\phi_{02}(t)\rangle$ contributes to the QB energy [see Fig. \ref{stablid}(d)]. This explains why the QB returns its ideal cyclic stage with period $T$ in the nonresonant case.

\section{Discussion and conclusions}\label{sec:Conclusion}
Our setup is two-level systems influenced by their radiative electromagnetic field propagating in a two-dimensional periodic structure as environments \cite{PhysRevLett.119.143602}. It can be realized by ultracold atoms held by optical lattice \cite{PhysRevLett.104.203603,Kri2018} or confined in photonic crystal \cite{PhysRevX.2.011014} and by transmon qubits in a circuit QED system \cite{PhysRevA.86.023837,Liu2017}. The system-environment bound state and its distinguished role in the static open-system dynamics have been observed in circuit QED \cite{Liu2017} and ultracold-atom \cite{Kri2018} systems. Floquet engineering has become a versatile tool in artificially synthesizing exotic quantum matters \cite{Zhang2017,Graphene}. This progress provides strong support to our scheme. It also indicates that the manipulation of the FBSs in the quasienergy spectrum via Floquet engineering is realizable in the state-of-the-art technique of quantum optics experiments.

In summary, we have investigated the decoherence dynamics of the charging-storage-discharging cyclic evolution of a QB-charger setup under the influence of the environments. It is found that, in sharp contrast to the Markovian approximate result where the QB asymptotically approaches deactivation, the QB can be kept alive in the non-Markovian dynamics. We have revealed that the mechanism behind this is the formation of two FBSs in the quasienergy spectrum of the total system consisting of the setup and the environments. In the resonant case, as long as the quasienergies of the two FBSs are near  degenerate, the QB would be reactivated to a near-ideal cyclic stage. In the nonresonant case, the QB is also reactivated by increasing the QB-charger coupling. Our result opens an avenue to beat the decoherence of QB and to build a lossless QB by Floquet engineering.

\section{Acknowledgments}
S.-Y.B. thanks Chong Chen and Wei Wu for fruitful discussions. This work is supported by the National Natural Science Foundation (Grants No. 11875150 and No. 11834005).

\bibliography{my}

\end{document}


\title{Supplemental material for ``Floquet engineering to reactivate a dissipative quantum battery''}
\author{Si-Yuan Bai}
\affiliation{School of Physical Science and Technology, Lanzhou University, Lanzhou 730000, China}
\author{Jun-Hong An}
\email{anjhong@lzu.edu.cn}
\affiliation{School of Physical Science and Technology, Lanzhou University, Lanzhou 730000, China}
\maketitle

\section{Ideal case}
The cyclic evolution of charging, storing, and discharging of the quantum battery (QB) is governed by the Hamiltonian
\begin{equation}
\hat{H}_0(t)=\sum_{l=\text{b,c}}\hbar\omega_{l}\hat{\sigma}_{l}^{\dagger}\hat{\sigma}_{l}+\hbar\kappa f(t)(\hat{\sigma}_\text{b}^{\dagger}\hat{\sigma}_\text{c}+\text{H.c.}),\label{eq:Ham-sys}
\end{equation}where $\hat{\sigma}_l=|g_l\rangle\langle e_l|$, with $|g\rangle$ and $|e\rangle$ being the ground and excited states of the two-level systems, are the transition operators of the QB and charger with frequency $\omega_l$, and $\kappa$ is their coupling strength. We redefine $\omega_0=(\omega_\text{c}+\omega_\text{b})/2$ and $\delta=(\omega_\text{c}-\omega_\text{b})/2$ for concise representation. The time-dependent $f(t)$ is
\begin{equation}
f(t)=\begin{cases}
1, ~ nT<t\le\tau_{c}+nT\\
0, ~ \tau_{c}+nT<t\le\tau_{c}+\tau_{s}+nT\\
1, ~ \tau_{c}+\tau_{s}+nT<t\le(n+1)T
\end{cases}\label{eq:f(t)}, ~n\in \mathbb{N},
\end{equation}
where $\tau_c$, $\tau_s$, and $\tau_d$ are the time of charging, storing, and discharging, respectively, and $T=\tau_c+\tau_s+\tau_d$ is the one-cycle period. Given the initial state $|\psi(0)\rangle=|g_\text{b},e_\text{c}\rangle$, the state at any time $|\psi(t)\rangle$ can be obtained by solving Schr\"{o}dinger equation governed by Eq. \eqref{eq:Ham-sys}. The performance of the QB is characterized by its mean energy $\mathcal{E}(t)=\langle\psi(t)|\hat{\sigma}_\text{b}^\dag\hat{\sigma}_\text{b}|\psi(t)\rangle$. The optimal working stage of the QB requires that the QB reaches its maximal energy at the end of each charging process, and empties its energy at the end of each discharging process, i.e., $|\psi(nT)\rangle=|\psi(0)\rangle$ or $\mathcal{E}(nT)=0$.

\textbf{Charging process}: Expanding the time-dependent state $|\psi(t)\rangle$ as
$|\psi(t)\rangle=\sum_{i=\text{b},\text{c}}c_{i}(t)\hat{\sigma}_{i}^{\dagger}|g_\text{b},g_\text{c}\rangle$, we can calculate that $c_\text{b}(t)$ and $c_\text{c}(t)$ satisfy
\begin{align}
i\dot{c}_\text{b}(t)= & \omega_\text{b}c_\text{b}(t)+\kappa c_\text{c}(t),\\
i\dot{c}_\text{c}(t)= & \omega_\text{c}c_\text{c}(t)+\kappa c_\text{b}(t),
\end{align}
under the initial condition $c_\text{b}(nT)=0$ and $c_\text{c}(nT)=1$. Their solutions read
\begin{align}
c_\text{b}(t) & =-i\frac{\kappa}{\Omega}\sin(\Omega t),\\
c_\text{c}(t) & =\cos(\Omega t)-i\frac{\delta}{\Omega}\sin(\Omega t),
\end{align}
where $\Omega=\sqrt{\kappa^{2}+\delta^{2}}$ and their common phase factor $\omega_0 t$ has been neglected. The energy of QB can be readily calculated
as $\mathcal{E}=\hbar\omega_\text{b}\kappa^2\sin^{2}(\Omega t)/\Omega^2$. The QB reaches its maximal value
\begin{equation}
\mathcal{E}_{{\rm max}}= \hbar\omega_\text{b}\kappa^2/\Omega^2,
\end{equation} when $t=(1/2+ n_{1})\pi/\Omega$. Thus we choose the time duration of the charging process as $\tau_c=(1/2+ n_{1})\pi/\Omega$. The state at the end of each charging process, i.e., $t=nT+\tau_c$, reads
\begin{equation}
|\psi(nT+\tau_{c})\rangle=\frac{1}{\Omega}[\kappa\hat{\sigma}_\text{b}^{\dagger}+\delta\hat{\sigma}_\text{c}^{\dagger}]|g_\text{b},g_\text{c}\rangle.\label{cahgd}
\end{equation}

\textbf{Storing process}: In the storing stage, the QB-charger coupling is switched off. Then the state evolves as
\begin{align}
|\psi(t)\rangle= & \frac{1}{\Omega}[\kappa e^{i2\delta(t-nT-\tau_{c})}\hat{\sigma}_\text{b}^{\dagger}+\delta\hat{\sigma}_\text{c}^{\dagger}]|g_\text{b},g_\text{c}\rangle.
\end{align}
At time $t=nT+\tau_c +\pi n_{2}/\delta$, the state can go back to Eq. \eqref{cahgd}. Thus we choose the time duration of the storing process as $\tau_s=\pi n_{2}/\delta$. The state at the end of each storing process, i.e., $t=nT+\tau_c+\tau_s$, reads
\begin{equation}
|\psi(nT+\tau_{c}+\tau_s)\rangle=\frac{1}{\Omega}[\kappa\hat{\sigma}_\text{b}^{\dagger}+\delta\hat{\sigma}_\text{c}^{\dagger}]|g_\text{b},g_\text{c}\rangle.\label{ddfdd}
\end{equation}

\textbf{Discharging process}: The QB-charger interaction is switched on again. The evolved state reads
\begin{eqnarray}
|\psi(t)\rangle&=&\hat{U}_0'(t-nT-\tau_c-\tau_s)|\psi(nT+\tau_{c}+\tau_s)\rangle\nonumber\\
&=&\hat{U}_0'(t-nT-\tau_c-\tau_s)|\psi(nT+\tau_{c})\rangle\nonumber\\
&=&\hat{U}_0'(t-\tau_s)|\psi(0)\rangle\nonumber\\
&=&\frac{1}{\Omega}\Big\{{\kappa}\sin[\Omega(t-\tau_s)]\hat{\sigma}^\dag_\text{b}+\{i\cos[\Omega(t-\tau_s)]\nonumber\\
&&+\delta\sin[\Omega(t-\tau_s)]\}\hat{\sigma}^\dag_\text{c}\Big\}|g_\text{b},g_\text{c}\rangle,
\end{eqnarray}where $\hat{U}_0'(t)=e^{{ -\frac{i}{\hbar}}\hat{H}'t}$ with $\hat{H}'=\sum_{l=\text{b,c}}\hbar\omega_{l}\hat{\sigma}_{l}^{\dagger}\hat{\sigma}_{l}+\hbar\kappa (\hat{\sigma}_\text{b}^{\dagger}\hat{\sigma}_\text{c}+\text{H.c.})$. When $t=\tau_{s}+\pi n_{3}/\Omega$, which defines a period $T=\tau_c+\tau_s+\tau_d$ with $\tau_d=(1/2+ n_{3})\pi/\Omega$, we have $|\psi(T)\rangle=|\psi(0)\rangle$.

\section{Floquet engineering}
Here we consider that the respective radiative electromagnetic field of the QB and charger as the environments are propagating in $N\times N$ 2D square-lattice. It can be realized by ultracold atoms held by optical lattice \cite{PhysRevLett.104.203603,Kri2018} or confined in photonic crystal \cite{PhysRevX.2.011014} and by transmon qubits in circuit QED system \cite{PhysRevA.86.023837,Liu2017}. The environmental Hamiltonian reads \cite{PhysRevA.96.043811}
\begin{equation}
\hat{H}_{\text{E}}=\hbar\sum_{l=\text{b,c}}  \sum_{\langle\mathbf{m},\mathbf{n}\rangle} [ \varpi \hat{b}^\dag_{l,\mathbf{n}}\hat{b}_{l,\mathbf{n}}-q(\hat{b}^\dag_{l,\mathbf{m}}\hat{b}_{l,\mathbf{n}}+\hat{b}^\dag_{l,\mathbf{n}}\hat{b}_{l,\mathbf{m}})],	\label{shmt}
\end{equation}
where $\langle \rangle$ means the summation is over all nearest neighbor sites, $\hat{b}_{l,\textbf{n}}$ are the annihilation operators of bosonic mode at \textbf{n}th site, $\varpi$ is their identical frequency, and $q$ is the nearest neighbour hopping rate.
The QB and charger reside in one local site $\mathbf{n}_0=(0,0)$ of the lattice. They couple to the bosonic modes in the lattice as
\begin{equation}
\hat{H}_\text{I}=\hbar g \sum_{l=\text{b,c}}(\hat{b}^\dag_{l,\mathbf{n}_0} \hat{\sigma}_l+\hat{b}_{l,\mathbf{n}_0} \hat{\sigma}^\dag_l) \label{sinter}
\end{equation} with $g$ being the coupling strength. Making the Fourier transformation to the operators $\hat{b}_{l,\mathbf{n}}=\frac{1}{N} \sum_{\mathbf{k}} \hat{b}_{l,\mathbf{k}} e^{i \mathbf{k}\cdot \mathbf{n}}$, where  $\mathbf{k}\equiv (k_x,k_y)$ are the momentum, we can rewrite Eqs. \eqref{shmt} and \eqref{sinter} into
\begin{eqnarray}
\hat{H}_\text{E}&=&\sum_{l=\text{b,c}}\sum_{\mathbf{k}}\hbar\omega_{\mathbf{k}}\hat{b}_{l,\mathbf{k}}^{\dagger}\hat{b}_{l,\mathbf{k}},\\
\hat{H}_\text{I}&=&\sum_{l=\text{b,c}}\sum_{\mathbf{k}}\hbar(g_{\mathbf{k}}\hat{b}_{l,\mathbf{k}}^{\dagger}\hat{\sigma}_l+\text{H.c.}),~\label{eq:H_interaction-1}
\end{eqnarray}
where $\omega_{\mathbf{k}}=\varpi-2q(\cos k_{x}+\cos k_{y})$ and $g_{\mathbf{k}}= g/N$.

\subsection{Resonant case}

We can develop a perturbation method to calculate the discrete quasienergies of the two FBSs. The Floquet equation reads
\begin{equation}
[\hat{H}_0(t)+\hat{H}_\text{E}+\lambda\hat{H}_\text{I}-i\hbar \partial_t]|\phi_{m,\alpha}(t)\rangle\rangle=\epsilon_{m,\alpha}|\phi_{m,\alpha}(t)\rangle\rangle,
\end{equation}where $\lambda$ characterizes the order of the perturbation and will be set to unit at the end of the derivation. The notation $|\rangle\rangle$ means that such states are in extended space and satisfy the following orthogonality \cite{PhysRevA.7.2203,Eckardt_2015}
\begin{eqnarray}
\langle\langle \phi_{m,\alpha}(t)|\phi_{n,\beta}(t)\rangle\rangle&\equiv&\frac{1}{T}\int_{0}^{T}dt\langle \phi_{m,\alpha}(t)|\phi_{n,\beta}(t)\rangle\nonumber
\\&=&\delta_{\alpha\beta}\delta_{mn}.
\end{eqnarray}The extended space is named Sambe space, which is made up of the usual Hilbert space and an extra temporal space. The label $m$ just characterizes the temporal space.
After rewriting $\epsilon_{m,\alpha}=\epsilon_{m,\alpha}^{(0)}+\lambda\epsilon_{m,\alpha}^{(1)}+\lambda^2\epsilon_{m,\alpha}^{(2)}$ and $|\phi_{m,\alpha}(t)\rangle=|\phi_{m,\alpha}^{(0)}(t)\rangle+\lambda|\phi_{m,\alpha}^{(1)}(t)\rangle$, we can obtain the terms in different power of $\lambda$ as
\begin{eqnarray}
&&[\hat{H}_{0}(t)+\hat{H}_\text{E}-i\hbar\partial_{t}]|\phi_{m,\beta}^{(0)}(t)\rangle\rangle=\epsilon_{m,\beta}^{(0)}|\phi_{m,\beta}^{(0)}(t)\rangle\rangle,\label{zerte}\\
&&[\hat{H}_{0}(t)+\hat{H}_\text{E}-i\hbar\partial_{t}]|\phi_{m,\beta}^{(1)}(t)\rangle\rangle+\hat{H}_\text{I}|\phi_{m,\beta}^{(0)}(t)\rangle\rangle=\epsilon_{m,\beta}^{(0)}\nonumber\\
&&~~~~~~~~~\times|\phi_{\beta,m}^{(1)}(t)\rangle\rangle+\epsilon_{m,\beta}^{(1)}|\phi_{m,\beta}^{(0)}(t)\rangle\rangle,\\
&&\hat{H}_\text{I}|\phi_{\beta,m}^{(1)}(t)\rangle\rangle=\epsilon_{\beta,m}^{(1)}|\phi_{\beta,m}^{(1)}(t)\rangle\rangle+\epsilon_{\beta,m}^{(2)}|\phi_{\beta,m}^{(0)}(t)\rangle\rangle.
\end{eqnarray}Then we have the first- and second-order corrections to the quasienergies of the Floquet bound states (FBSs)
\begin{eqnarray}
\epsilon_{m,\beta}^{(1)}&=&\langle\langle \phi_{m,\beta}^{(0)}(t)|\hat{H}_\text{I}|\phi_{m,\beta}^{(0)}(t)\rangle\rangle,\\
\epsilon_{m,\beta}^{(2)}&= & \sum_{n,\gamma}\frac{|\langle\langle \phi_{n,\gamma}^{(0)}(t)|\hat{H}_\text{I}|\phi_{m,\beta}^{(0)}(t)\rangle\rangle|^{2}}{\epsilon_{m,\beta}^{(0)}-\epsilon_{n,\gamma}^{(0)}}.\label{2-order-energy}
\end{eqnarray}
In the following, we calculate the corrections in different orders of perturbation.

The zeroth-order equation \eqref{zerte} is equivalent to
\begin{equation}
\hat{U}_0(T)|\phi^{(0)}_{m,\alpha}(0)\rangle=e^{{-i\over\hbar}\epsilon^{(0)}_{m,\alpha}T}|\phi^{(0)}_{m,\alpha}(0)\rangle,\label{uttt}
\end{equation}where $\hat{U}_0(T)=\mathcal{T}e^{{-i\over\hbar}\int_0^T[\hat{H}_0(t)+\hat{H}_\text{E}]dt}=e^{{-i\over\hbar}\hat{H}_{0,\text{eff}}T}$ with \begin{eqnarray}
\hat{H}_{0,\text{eff}}=\hbar\omega_0\sum_{l=\text{b,c}}\hat{\sigma}_{l}^{\dagger}\hat{\sigma}_{l}+{2\hbar\kappa\over 3}(\hat{\sigma}_\text{b}^{\dagger}\hat{\sigma}_\text{c}+\text{H.c.})+\hat{H}_\text{E}.~~~
\end{eqnarray}Here we have assumed $\tau_c=\tau_s=\tau_d=\pi/(2\kappa)$, which satisfy the optimal working condition derived in the last section. The solution of Eq. \eqref{uttt} in the single-excitation space can be readily obtained. Then applying $\hat{U}_0(t)$ on $|\phi^{(0)}_\alpha(0)\rangle$, we obtain
\begin{eqnarray}
\epsilon^{(0)}_{m,l,{\bf k}}&=&\hbar\omega_{\bf{k}}+m\hbar\omega_T,\label{eigen-value-1}\\
|\phi^{(0)}_{m,l,{\bf k}}(t)\rangle\rangle&=&e^{im\omega_Tt}\hat{b}_{l,{\bf k}}^\dag|{\O}\rangle,~~\label{eigen-vector-1}\\
\epsilon^{(0)}_{m,+}&=&\hbar\omega_0+(m+{1\over 2})\hbar\omega_T,\label{eigen-value-2}\\
|\phi^{(0)}_{m,+}(t)\rangle\rangle&=&y(t)e^{im\omega_Tt}\frac{\hat{\sigma}^\dag_\text{b}+\hat{\sigma}^\dag_\text{c}}{\sqrt{2}}|{\O}\rangle,~~\label{eigen-vector-2}\\
\epsilon^{(0)}_{m,-}&=&\hbar\omega_0+(m-{1\over 2})\hbar\omega_T,\label{eigen-value-3}\\
|\phi^{(0)}_{m,-}(t)\rangle\rangle&=&y^*(t)e^{im\omega_Tt}\frac{\hat{\sigma}^\dag_\text{b}-\hat{\sigma}^\dag_\text{c}}{\sqrt{2}}|{\O}\rangle,~~\label{eigen-vector-3}
\end{eqnarray}
where $l=\rm{b},\rm{c}$, we have used $\omega_{\text{b},\mathbf{k}}=\omega_{\text{c},\mathbf{k}}\equiv  \omega_{\mathbf{k}}$ and \begin{equation}
y(t)=\begin{cases}
e^{-i\frac{\kappa}{3}t}, & nT<t\le nT+\frac{T}{3},\\
e^{-i\frac{\kappa}{3}(T-2t)}, &nT+ \frac{T}{3}<t\le nT+\frac{2T}{3},\\
e^{-i\frac{\kappa}{3}(t-T)}, & nT+\frac{2T}{3}<t\le (n+1)T.
\end{cases}
\end{equation}

We can readily check that the first-order corrections are zero
\begin{equation}
\epsilon_{m,\beta}^{(1)}=\langle\langle \phi_{m,\beta}^{(0)}(t)|\hat{H}_\text{I}|\phi_{m,\beta}^{(0)}(t)\rangle\rangle=0.
\end{equation}
The second-order corrections of $\epsilon_{-1,+}^{(0)}$ and $\epsilon_{0,-}^{(0)}$ are
\begin{eqnarray}
\epsilon_{-1,+}^{(2)}&= & 2\sum_{\mathbf{k},n}\frac{|\frac{\hbar g_{\mathbf{k}}}{\sqrt{2}}\frac{1}{T}\int_{0}^{T}dt e^{-i(n+1)\omega_{T}t}y(t)|^{2}}{\epsilon_{-1,+}^{(0)}-\hbar\omega_{\mathbf{k}}-n\hbar\omega_{T}}\nonumber\\
&=&\sum_{\mathbf{k},n}\frac{\hbar^2 g_{\mathbf{k}}^{2}|f_{n}|^{2}}{\epsilon_{-1,+}^{(0)}-\hbar\omega_{\mathbf{k}}-(n-1)\hbar\omega_{T}},\label{sedpp}\\
\epsilon_{0,-}^{(2)}&= & 2\sum_{\mathbf{k},n}\frac{|\frac{\hbar g_{\mathbf{k}}}{\sqrt{2}}\frac{1}{T}\int_{0}^{T}dte^{-in\omega_{T}t}y^{*}(t)|^{2}}{\epsilon_{0,-}^{(0)}-\hbar\omega_{\mathbf{k}}-n\hbar\omega_{T}}\nonumber\\
&=&\sum_{\mathbf{k},n}\frac{\hbar^2 g_{\mathbf{k}}^{2}|f_{n}|^{2}}{\epsilon_{0,-}^{(0)}-\hbar\omega_{\mathbf{k}}+n\hbar\omega_{T}},\label{sedmm}
\end{eqnarray}
where $f_{n}=\frac{1}{T}\int_{0}^{T}dte^{-in\omega_{T}t}y(t)$. Note that, although the two quasienergies of $\epsilon_{-1,+}^{(0)}$ and $\epsilon_{0,-}^{(0)}$ are degenerate, the non-degenerate perturbation theory we developed is still applicable because the perturbation $\hat{H}_\text{I}$ does not cause mixture of the two FBSs $|\phi^{(0)}_{-1,+}(t)\rangle\rangle$ and $|\phi^{(0)}_{0,+}(t)\rangle\rangle$.

\begin{figure}
\centering
\includegraphics[width=1\columnwidth]{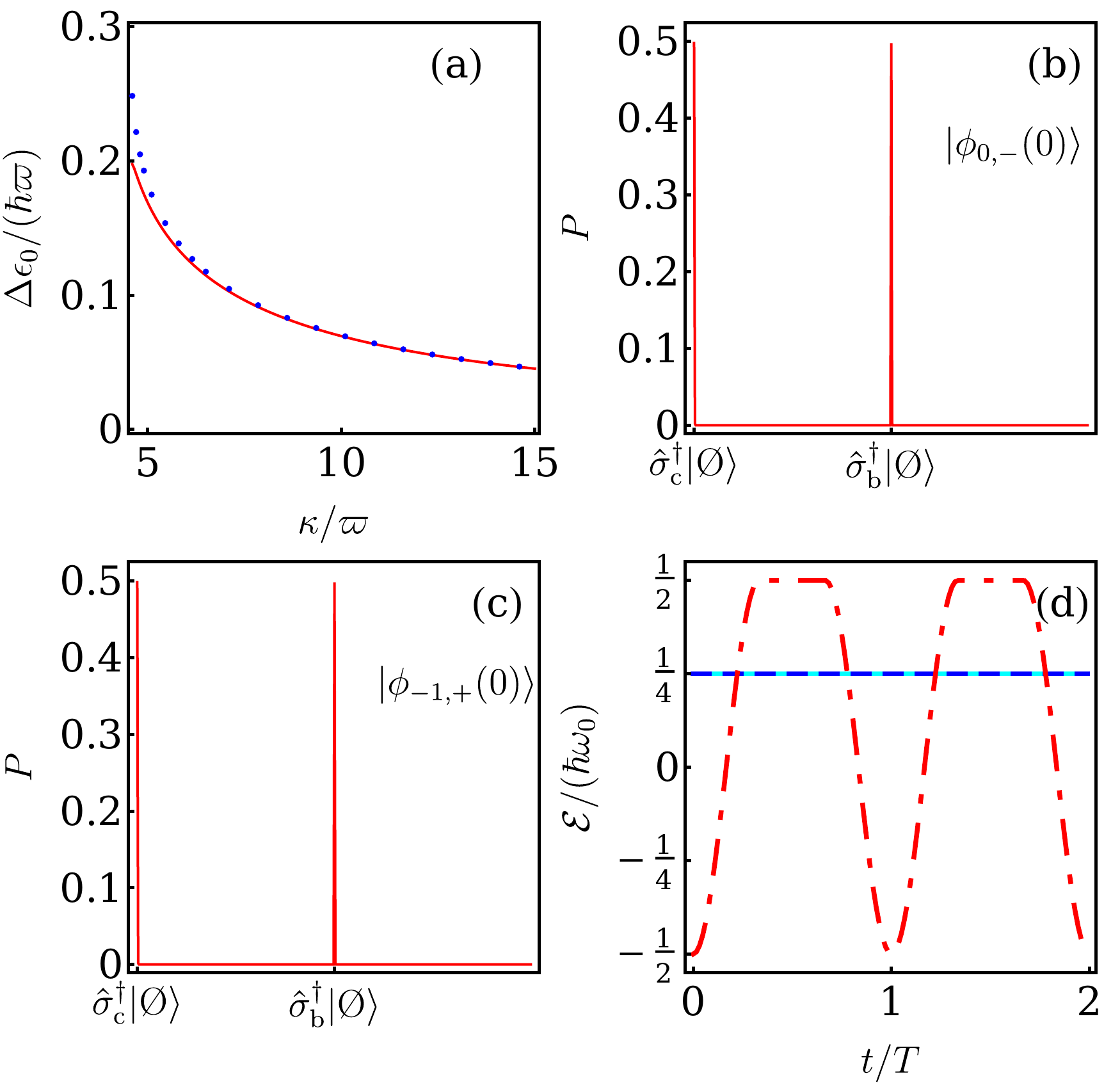}\caption{(a) Quasienergies of the two FBSs calculated via numerically solving the Floquet equation (red solid lines) and via analytically solving the perturbation equations \eqref{sedpp} and \eqref{sedmm} (blue dots). (b) and (c) Probability distributions of $|\phi_{01}(0)\rangle$ and $|\phi_{02}(0)\rangle$ calculated via numerically solving the Floquet equation. They validate Eqs. (\ref{sedpp}) and (\ref{sedmm}). (d) Diagonal terms for $j=j'=(-1,+)$ (cyan solid line) and $(0,-)$ (blue dashed line), and interference terms for $j\neq j'$ (red dotdashed line) of Eq. \eqref{stdav} when $\kappa=15\varpi$ analytically evaluated from the perturbation method. It matches with the numerical results in Fig. 4(b) in the main text. Other parameter values are the same as the ones in Figs. 4(a) and 4(b) in the main text.}\label{ffgsm}
\end{figure}
Using the parameter values of Fig. 4(a) in the main text, we plot in Fig. \ref{ffgsm}(a) the quasienergy difference of the two FBSs via numerically solving the Floquet equation and via analytically evaluating the above perturbation corrections. It shows clearly that the perturbation theory works well in the large $\kappa$ parameter regime. Then we can roughly evaluate the mean energy of the QB by the perturbation quasienergies in Eqs. (\ref{sedpp}, \ref{sedmm}). Figures \ref{ffgsm}(b) and \ref{ffgsm}(c) confirm the correctness of the perturbation results in Eqs. \eqref{eigen-vector-2} and \eqref{eigen-vector-3} at $t=0$, which readily induce $c_1=-c_2=1/\sqrt{2}$ according to the initial state $|\Psi(0)\rangle$. We also can obtain the contributions of the two FBSs to the energy of the QB using Eqs. \eqref{eigen-vector-2} and \eqref{eigen-vector-3}. The diagonal terms read $\langle \phi^{(0)}_{-1,+}(t)|\hat{\sigma}_\text{b}^{\dagger}\hat{\sigma}_\text{b}|\phi^{(0)}_{-1,+}(t)\rangle=\langle \phi^{(0)}_{0,-}(t)|\hat{\sigma}_\text{b}^{\dagger}\hat{\sigma}_\text{b}|\phi^{(0)}_{0,-}(t)\rangle=\frac{1}{2}$, and the interference terms read  \begin{eqnarray}
&&\text{Re}[e^{-\frac{i}{\hbar}\Delta\epsilon_0 t}\langle \phi^{(0)}_{0,-}(t)|\hat{\sigma}_\text{b}^{\dagger}\hat{\sigma}_\text{b}|\phi^{(0)}_{-1,+}(t)\rangle]\nonumber\\
\approx ~&&\cos(\frac{\Delta\epsilon_0}{\hbar}t)\text{Re}[\frac{y^2(t)e^{-i\omega_T t}}{2}].
\end{eqnarray} Then from
\begin{equation}
{\mathcal{E}(\infty)\over\hbar\omega_0}=\sum_{jj'=1}^Mc_jc_{j'}^*e^{{-i\over\hbar}(\epsilon_{0j}-\epsilon_{0j'})t}\langle \phi_{0j'}(t)|\hat{\sigma}_\text{b}^\dag\hat{\sigma}_\text{b}|\phi_{0j}(t)\rangle, ~~~~~~~~\label{stdav}
\end{equation}we have
\begin{eqnarray}
{\mathcal{E}(\infty)\over\hbar\omega_0}= \frac{1}{2}\Big\{1-\cos(\frac{\Delta\epsilon_0}{\hbar}t)\text{Re}[y^2(t)e^{-i\omega_T t}]\Big\}.\label{stdaval}
\end{eqnarray}

With increasing $\kappa$, the gap between two quasienergies of the two FBSs tends to zero and the oscillation is slowed down. These analytical components are shown in Fig. \ref{ffgsm}(d) with the same parameter values in Fig. 4(b) of main text, which match well with the numerical results.


\subsection{Nonresonant case}

\begin{figure}
\centering
\includegraphics[width=1\columnwidth]{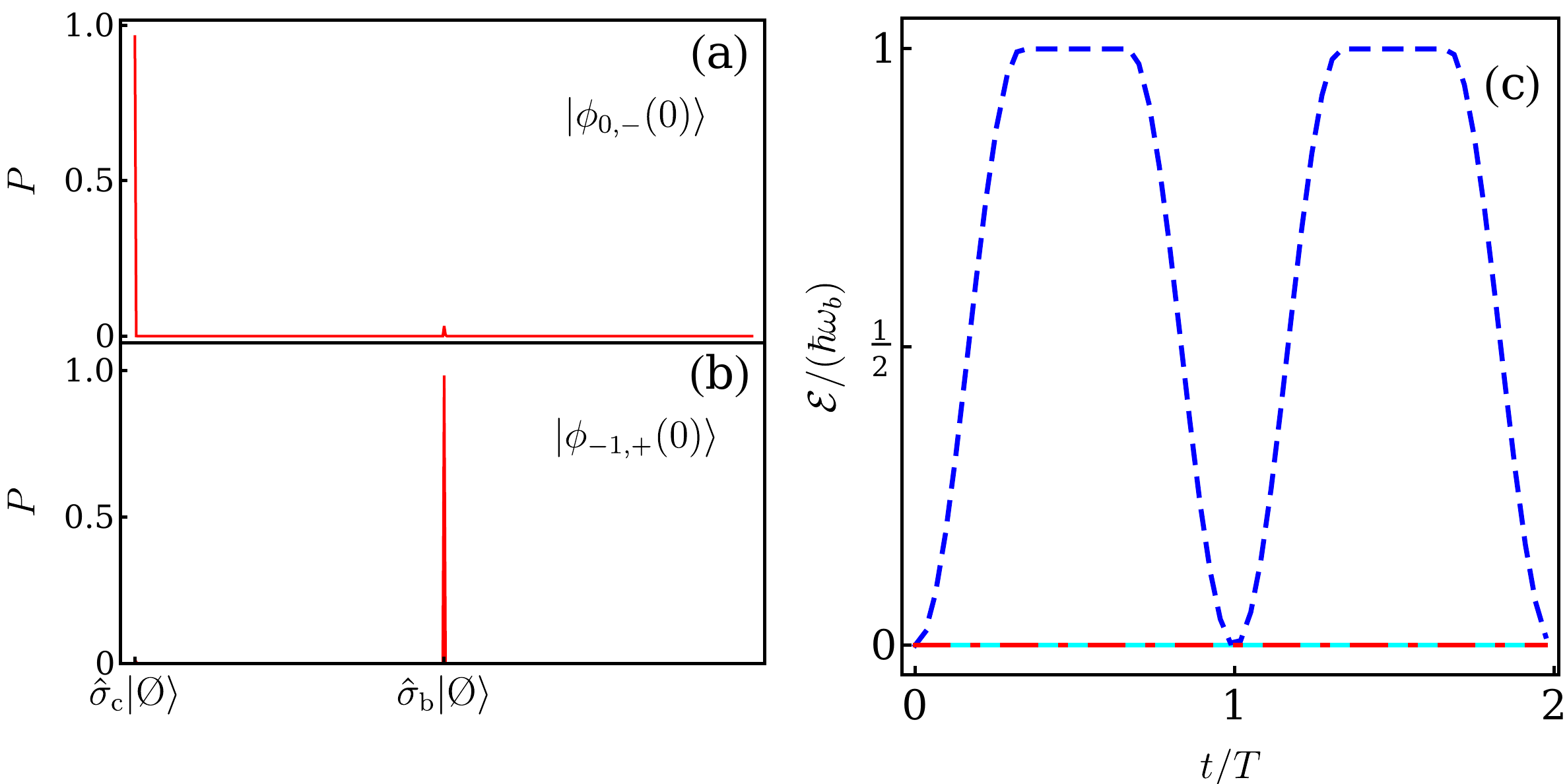}\caption{(a) and (b) Probability distribution of the two FBSs calculated via numerically solving the Floquet equation, which matches with Eqs. \eqref{eigenv-2} and \eqref{eigenv-3}. (c) Diagonal terms for $j=j'=(-1,+)$ (cyan solid line) and $(0,-)$ (blue dashed line), and interference terms for $j\neq j'$ (red dotdashed line) of Eq. \eqref{stdav} when $\kappa=15\varpi$ analytically evaluated from the perturbation method. Other parameter values are the same as the ones in Fig. 4 (c) and (d) in the main text.}\label{fig:non-res}
\end{figure}
If a frequency detuning $\delta$ between the QB and charger is present, the degeneracy of $\epsilon_{-1,+}^{(0)}$ and $\epsilon_{0,-}^{(0)}$ would be broken. We can use the degenerate perturbation theory to re-derive the zeroth-order FBSs. The Hamiltonian in this case reads
\begin{eqnarray}
\hat{H}_{\delta}(t)=&&\hbar\omega_{0}\sum_l\hat{\sigma}_{l}^{\dagger}\hat{\sigma}_{l}+\hbar f(t)(\hat{\sigma}_{b}^{\dagger}\hat{\sigma}_{c}+\hat{\sigma}_{c}^{\dagger}\hat{\sigma}_{b})\nonumber\\
&&+\hbar \delta(\hat{\sigma}_{\rm{c}}^{\dagger}\hat{\sigma}_{\rm{c}}-\hat{\sigma}_{\rm{b}}^{\dagger}\hat{\sigma}_{\rm{b}})+\hat{H}_\text{E}. \label{non-res-eff-H}
\end{eqnarray} Treating the $\delta$ term as perturbation, we can evaluate the first-order correction to $\epsilon_{-1,+}^{(0)}$ and $\epsilon_{0,-}^{(0)}$ in Eqs. \eqref{eigen-value-2} and \eqref{eigen-value-3} via
\begin{eqnarray}
 	\det [\hbar \delta \left(
                       \begin{array}{cc}
                         1 & 0 \\
                         0 & -1 \\
                       \end{array}
                     \right)
 -\epsilon^{(0)}\mathbb{I}]=0.
 \end{eqnarray}
We thus have
\begin{eqnarray}
\epsilon^{(0)}_{m,+}&=&\hbar\omega_0+(m+{1\over 2})\hbar\omega_T+\hbar\delta,\label{eigen2}\\
|\phi^{(0)}_{m,+}(0)\rangle\rangle&=&\hat{\sigma}^\dag_\text{b}|{\O}\rangle,~~\label{eigenv-2}\\
\epsilon^{(0)}_{m,-}&=&\hbar\omega_0+(m-{1\over 2})\hbar\omega_T-\hbar\delta,\label{eigen3}\\
|\phi^{(0)}_{m,-}(0)\rangle\rangle&=&\hat{\sigma}^\dag_\text{c}|{\O}\rangle.~~\label{eigenv-3}
\end{eqnarray}
It can be seen that the two quasienergy difference $\Delta\epsilon_0=|\epsilon_{-1,+}-\epsilon_{0,-}|$ cannot be zero anymore. In this case we still have chance to make the QB return to its near ideal stage. Applying the evolution $\hat{U}_0(t)$ on Eqs. \eqref{eigenv-2} and \eqref{eigenv-3}, we have $|\phi^{(0)}_{-1,+}(t)\rangle\rangle$ and $|\phi^{(0)}_{0,-}(t)\rangle\rangle$. Then we can roughly evaluate the energy of the QB. The initial condition induces $c_{-1,+}=\langle\phi^{(0)}_{-1,+}(0)|\Psi(0)\rangle=0$ and $c_{0,-}=\langle\phi^{(0)}_{0,-}(0)|\Psi(0)\rangle=1$, which is confirmed by the numerical results in Figs. \ref{fig:non-res}(a) and \ref{fig:non-res}(b). Therefore, only the $|\phi^{(0)}_{0,-}(t)\rangle\rangle$ has contribution to the energy of the QB. It reads
\begin{eqnarray}
{\mathcal{E}(\infty)\over\hbar\omega_0}=\langle \phi^{(0)}_{0,-}(t)|\hat{\sigma}_\text{b}^\dag\hat{\sigma}_\text{b}|\phi^{(0)}_{0,-}(t)\rangle,
\end{eqnarray}which returns to the $T$-periodic. Therefore, the QB goes back to the ideal working stage.

These analytical components of the two diagonal terms and the interference term of Eq. \eqref{stdav} are shown in Fig. \ref{fig:non-res}(c) with same parameters in Fig. 4(d) of the main text. Matching well the numerical results in Fig. 4(d) of the main text, it shows clearly that only one of the two diagonal terms dominates the energy. This also explain why the QB returns its ideal stage in the nonresonant case.

\bibliography{my}